\documentclass[12pt]{iopart}
\usepackage{amssymb}
\usepackage{nicefrac}
\usepackage{graphicx}

\begin{document}

\title{Towards resolution of the Fermi surface in underdoped high-$T_{\rm c}$ superconductors}
\author{Suchitra~E.~Sebastian$^1$, Neil~Harrison$^2$, Gilbert~G.~Lonzarich$^1$}
 
\address{$1$ Cavendish Laboratory, Cambridge University, JJ Thomson Avenue, Cambridge CB3~0HE, U.K}
\address{$2$ National High Magnetic Field Laboratory, Los Alamos National Laboratory, MS E536, Los Alamos, New Mexico 87545, U.S.A.}

\ead{suchitra@phy.cam.ac.uk}
\date{\today}

\begin{abstract}
 
We survey recent experimental results including quantum oscillations and complementary measurements probing the electronic structure of underdoped cuprates, and theoretical proposals to explain them. We discuss quantum oscillations measured at high magnetic fields in the underdoped cuprates that reveal a small Fermi surface section comprising quasiparticles that obey Fermi–Dirac statistics, unaccompanied by other states of comparable thermodynamic mass at the Fermi level. The location of the observed Fermi surface section at the nodes is indicated by a body of evidence including the collapse in Fermi velocity measured by quantum oscillations, which is found to be associated with the nodal density of states observed in angular resolved photoemission, the persistence of quantum oscillations down to low fields in the vortex state, the small value of density of states from heat capacity and the multiple frequency quantum oscillation pattern consistent with nodal magnetic breakdown of bilayer-split pockets. A nodal Fermi surface pocket is further consistent with the observation of a density of states at the Fermi level concentrated at the nodes in photoemission experiments, and the antinodal pseudogap observed by photoemission, optical conductivity, nuclear magnetic resonance (NMR) Knight shift, as well as other complementary diffraction, transport and thermodynamic measurements. One of the possibilities considered is that the small Fermi surface pockets observed at high magnetic fields can be understood in terms of Fermi surface reconstruction by a form of small wavevector charge order, observed over long lengthscales in experiments such as nuclear magnetic resonance and x-ray scattering, potentially accompanied by an additional mechanism to gap the antinodal density of states.

\tableofcontents
\end{abstract}
 
\maketitle


\section{Introduction}
\begin{figure}[htbp!]
\centering
\includegraphics[width=0.8\textwidth]{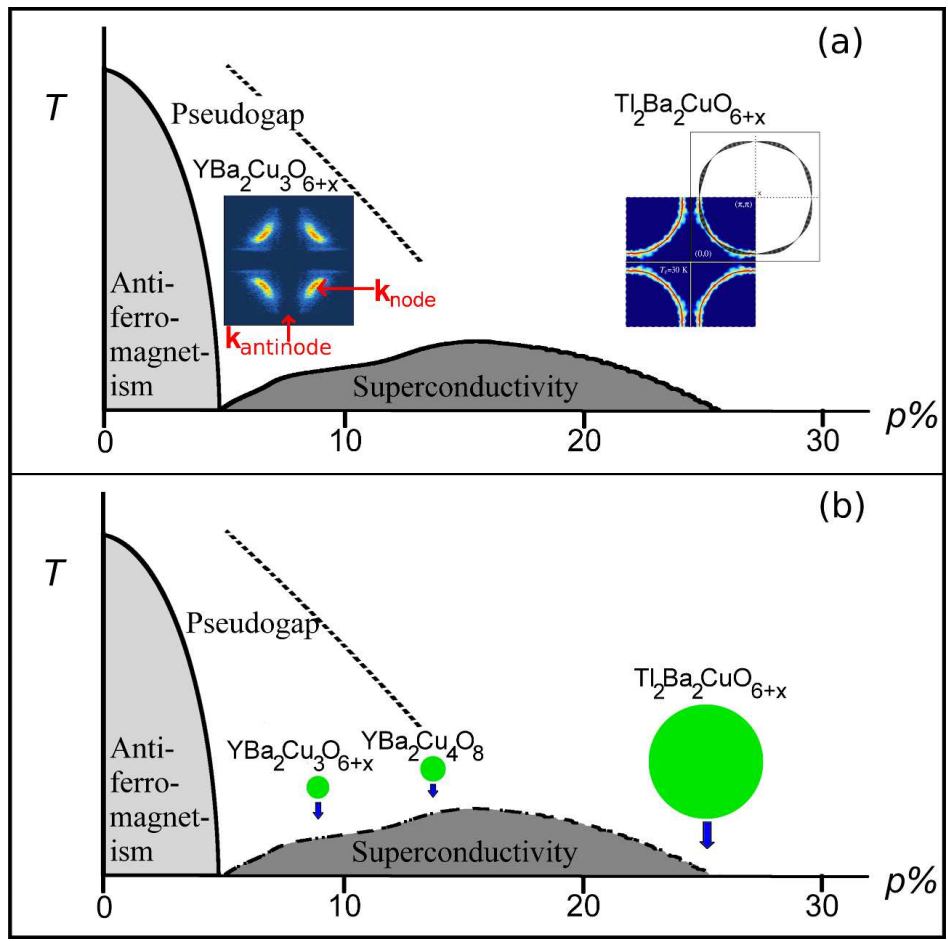}
\caption{(a) Schematic of the high-$T_{\rm c}$ cuprate phase diagram as a function of nominal hole doping ($p$) denoting electronic structure on the overdoped and underdoped side. Electronic structure on the overdoped side measured by photoemission experiments at $B = 0$ in overdoped Tl$_2$Ba$_2$CuO$_{6+\rm x}$, reproduced with permission from ref.~\cite{peets1}, overlaid with electronic structure measured by angular magnetoresistance experiments, reproduced with permission from ref.~\cite{hussey1}. Electronic structure on the underdoped side measured by photoemission experiments at $B=0$ in underdoped YBa$_2$Cu$_3$O$_{\rm {6+x}}$, reproduced with permission from ref.~\cite{hossain1}. For YBa$_2$Cu$_3$O$_{\rm {6+x}}$, the relation between Oxygen doping ($x$) and hole concentration ($p$) is given by $p \approx 18 x \%$~\cite{liang1}. Superconducting, antiferromagnetic, and pseudogap regions are indicated. (b) Schematic of the high-$T_{\rm c}$ cuprate phase diagram as a function of nominal hole doping ($p$) denoting electronic structure on the overdoped and underdoped side, measured by quantum oscillation experiments at high magnetic fields. The superconducting region at zero magnetic field is schematically represented by a dashed boundary. Circles represent the relative size of Fermi surface orbits observed in quantum oscillation experiments ~\cite{doiron1,yelland1,bangura1,vignolle1}.}
\label{phasediagram}
\end{figure}

The normal state out of which high $T_{\rm c}$ superconductivity develops in the cuprates has proved challenging to understand~\cite{bednorz1,orenstein1}. Yet its resolution may hold the key to understanding the origin of unconventional pairing in these materials~\cite{anderson1}. Angle-resolved photoemission spectroscopy (ARPES) measurements~\cite{damascelli1}, performed at zero magnetic field and elevated temperatures to suppress superconductivity and uncover the normal state in the cuprates, reveals the large Fermi surface in the overdoped regime~\cite{plate1,peets1,hussey1} ($p>$~20~\% in figure~\ref{phasediagram}) to be transformed into small disconnected regions called `Fermi arcs'  located near the nodal locations in the Brillouin zone (${\bf k}_{\rm node} = [\pm \pi/2, \pm \pi/2]$, shown in figure~\ref{phasediagram}a). A pseudogap of size $\approx$ 50~meV of not fully understood origin opens at the antinodal locations in the Brillouin zone (${\bf k}_{\rm antinode} = [0, \pm \pi/2], [\pm \pi/2, 0]$) in this so called `normal state' of the underdoped regime ($p<$~20~\%)~\cite{shen1, valla1,hossain1,meng1,king1,lee1,kanigel1}. The `nodal' and `antinodal' regions of the Brillouin zone are thus termed due to the corresponding minimally and maximally gapped locations of the d-wave superconducting gap function respectively.

\begin{figure}[htbp!]
\centering
\includegraphics[width=0.65\textwidth]{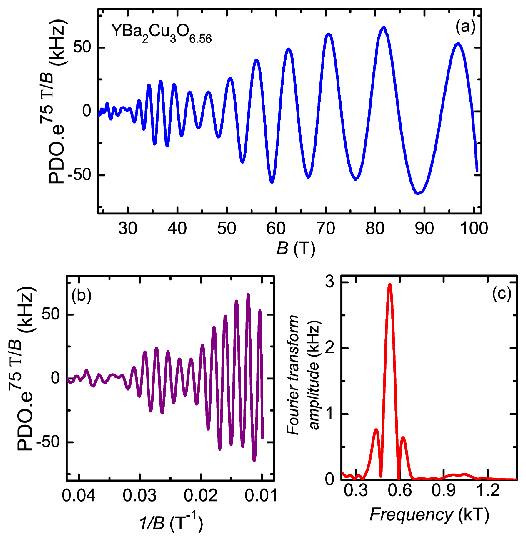}
\caption{(a) Quantum oscillations in underdoped YBa$_2$Cu$_3$O$_{\rm {6+x}}$ with $x=0.56$ (or $p \approx 10\%$) measured by the contactless resistivity technique (resonant frequency shift measured by a proximity detector oscillator denoted as `PDO') over a broad range of field extending to $\approx 101$~T (blue line) at $T\approx 1.5$~K, from ref.~\cite{sebastian8}. For visual clarity, the data has been multiplied by an exponential factor. On theoretical and experimental grounds it is expected that high-field magneto-oscillatory effects can be described in terms of a superposition of amplitude-modulated components periodic in inverse magnetic field ($1/B$).  These components are usually searched for in the first instance by examining not only the raw data as a function of $1/B$ (purple line in (b)), but also the Fourier decomposition of the data as a function of the conjugate frequency in $1/B$. Shown in (c) is the Fourier transform amplitude spectrum (in red) with a Blackman window function multiplying the raw data to attenuate spurious frequency side lobes arising from data end effects.
\\[8pt]
The identification of the underlying periodic components is in general made challenging by the field dependent modulation of each component, including windowing of the data train, which limits frequency resolution and accurate determinations of the amplitudes and phases of the individual components.  If the number of components and the mathematical forms of the amplitude modulations are known from additional empirical and theoretical considerations, curve fitting algorithms may be used to extract more accurate frequency, amplitude and phase information than is readily possible by conventional Fourier decomposition alone.  Curve fitting techniques normally follow after careful examinations of Fourier spectra, which are model free and (when both real and imaginary components are retained) complete in the sense that they can be used to rigorously reconstruct the original raw data versus magnetic field. 
\\[8pt]
The Fourier spectrum in (c) suggests the presence of a minimum of three frequency components corresponding to the central peak at $\approx 532$~T, flanked by side peaks at $\approx 440$ and $\approx 620$~T respectively, together with their (weaker) harmonics. Additional empirical and theoretical information regarding the form of the Fermi surface would be required in order to extract improved values of frequency, amplitude, and phase of the various components by curve fitting techniques. We note that hypotheses regarding the form of the Fermi surface have been used to fit the measured magneto-oscillatory data in underdoped YBa$_2$Cu$_3$O$_{\rm {6+x}}$, for example in refs.~\cite{sebastian8,ramshaw1}.
}
\label{100T}
\end{figure}

More recently, applied magnetic fields extending over a broad range from $\approx$~22~T to 101~T (figure~\ref{100T})~\cite{sebastian8} have been used to weaken superconductivity and reveal properties of the underlying normal state by means of quantum oscillations, uncovering a Fermi surface comprising small sections (figure~\ref{phasediagram}b) ~\cite{sebastian8,doiron1,yelland1,bangura1,sebastian3,leboeuf1,audouard1,sebastian6,sebastian2,sebastian4,singleton1,ramshaw1,sebastian5,riggs1,sebastian7,laliberte1,sebastian1,vignolle2} and associated with a low value of linear coefficient of heat capacity~\cite{riggs1}. The transformation of the large Fermi surface observed in overdoped Tl$_2$Ba$_2$CuO$_{6+\rm x}$~(figure~\ref{phasediagram}b, \cite{vignolle1}) and predicted from band structure calculations~\cite{andersen1} to the observed small size Fermi surface sections in underdoped cuprates is striking. Also remarkable is the degree to which the quasiparticles responsible for the observed quantum oscillations are governed by Fermi-Dirac statistics within the underdoped region~\cite{sebastian2}. The dramatic change in Fermi surface as compared with the overdoped side is accompanied by a negative sign of Hall~\cite{leboeuf1} and Seebeck~\cite{laliberte1} coefficients (in YBa$_2$Cu$_3$O$_{6+x}$ and YBa$_2$Cu$_4$O$_8$), seemingly characteristic of negatively charged quasiparticles, instead of a positive sign as expected for the large hole Fermi surface on the overdoped side.

Photoemission, optical conductivity and other experiments point to the existence of gapless excitations chiefly in the vicinity of the nodes. The finding of a predominantly single carrier Fermi surface (with multiple oscillatory components) in underdoped YBa$_2$Cu$_3$O$_{\rm {6+x}}$ measured in an extended set of quantum oscillation experiments~\cite{sebastian7} and the measured low value of linear coefficient of heat capacity~\cite{riggs1} pose additional constraints on the electronic structure. On considering alternatives for the location of the single carrier Fermi surface pocket at the antinodal or the nodal region of the Fermi surface, evidence points to the location of the Fermi surface pocket near the nodes of the Brillouin zone. Evidence includes (i) an antinodal pseudogap of $\approx$~50 meV ~\cite{norman1} observed by angular resolved photoemission~\cite{damascelli1}, optical conductivity~\cite{basov1}, and other experiments at zero magnetic field~\cite{alloul1}, which is unlikely to be suppressed in magnetic fields~\cite{tranquada1} as low as 22~T at which quantum oscillations begin to be observed (section 2.5.1), (ii) a collapse in Fermi velocity observed both by quantum oscillations and nodal angular resolved photoemission (section 2.5.2), (iii) the persistence of quantum oscillations down to low magnetic fields in the vortex solid regime (section 2.5.3), and (iv) the quantum oscillation waveform characteristic of magnetic breakdown of a nodal bilayer-split Fermi surface pocket (section 2.5.5).

We review existing models, and discuss in more detail proposals for the normal state electronic structure that have the potential to reconcile complementary experiments such as photoemission, heat capacity, and quantum oscillations. A biaxial form of small $\bf Q$ charge order with ordering wavevector similar to that observed in high field nuclear magnetic resonance and zero field resonant x-ray scattering (RXS) experiments is proposed to reconstruct the Fermi surface at low dopings. This would yield a nodal Fermi surface pocket consistent with the nodal density-of-states at the Fermi level observed by photoemission as `Fermi arcs' in zero field, and which would give rise to the experimentally observed negative Hall and Seebeck coefficients and low value of linear coefficient of heat capacity at high magnetic fields. The gapping of the antinodal density of states in such a picture could arise either from a distinct pseudogap origin, or the density wave responsible for the reconstruction of the Fermi surface into closed nodal Fermi surface pockets.

\section{Review of the experimental evidence}
We begin by reviewing key experimental properties relating to the electronic structure of the underdoped cuprates observed by quantum oscillation measurements, and relate these findings to observations made by photoemission, heat capacity, optical conductivity and other techniques.

\subsection{Fermi Dirac statistics}
\begin{figure}[htbp!]
\centering
\includegraphics[width=0.95\textwidth]{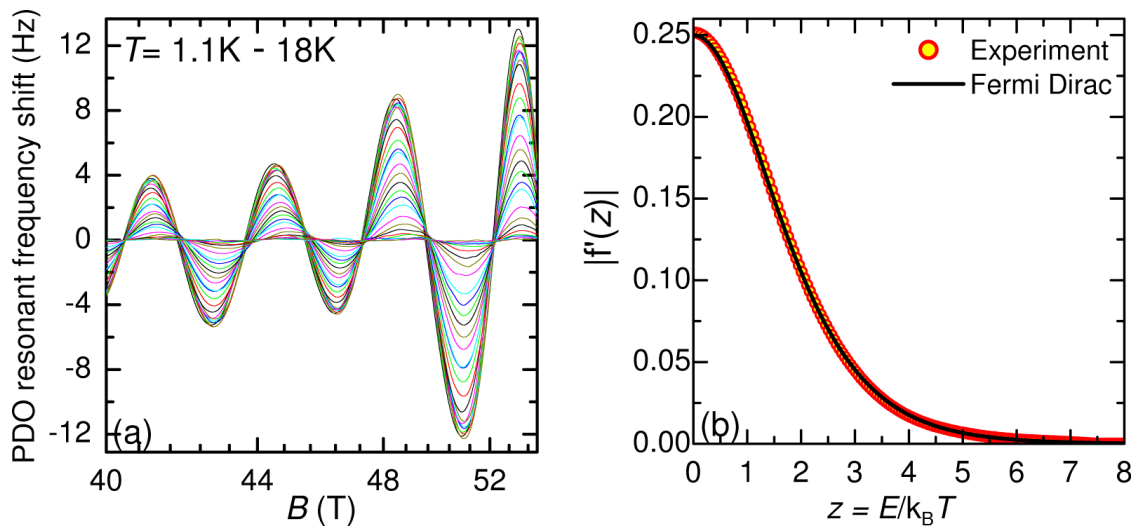}
\caption{(a) Measured quantum oscillations in the contactless resistivity (PDO resonant frequency shift) in underdoped YBa$_2$Cu$_3$O$_{6+x}$ with $x=$0.56 (or $p\approx$~10\%) shown over a broad range of temperatures 1.1~K~$\lessapprox T\lessapprox$~18~K, from ref.~\cite{sebastian2}. (b) Comparison of the derivative of the Fermi-Dirac distribution versus $E/k_{\rm B}T$ with that inferred from the temperature dependence of the amplitude of the quantum oscillations from 0.1 K to 18 K, showing excellent agreement. Here $f'(z)=\partial{f(z)}/\partial{z}$, where $f(z)$ is the average occupation number of quasiparticle states of excitation energy $E=k_{\rm B}Tz$, e.g. $|{f}^\prime_{\rm FD}(z)|=1/(2+2\cosh z)$ for the Fermi Dirac quasiparticle distribution.}
\label{fermidirac}
\end{figure}

The statistics governing the quasiparticles that execute cyclotron motion in quantizing magnetic fields in the underdoped cuprates has been determined by a careful study of the temperature-dependence of the quantum oscillation amplitude~\cite{sebastian2}. An inverse Fourier transform of the measured quantum oscillation amplitude in YBa$_2$Cu$_3$O$_{\rm{6+x}}$ (with respect to the dimensionless parameter $\eta=2\pi k_{\rm B}Tm^\ast/\hbar eB$) over a broad range in temperature 100~mK~$\lessapprox T\lessapprox$~18~K (figure~\ref{fermidirac}a)~\cite{sebastian2}, yields the quasiparticle statistical distribution (figure~\ref{fermidirac}b). On comparison with the expected statistical distribution for Fermi-Dirac quasiparticles, excellent correspondence with experiment is seen (figure~\ref{fermidirac}b~\cite{sebastian2}), establishing the low energy quasiparticles observed by quantum oscillations in YBa$_2$Cu$_3$O$_{6+x}$ at low temperatures and high magnetic fields to be Fermi liquid-like in nature. The observation of Fermi Dirac statistics is notable given the finding in complementary experiments such as photoemission of a vanishing quasiparticle residue at the nodes for low dopings in YBa$_2$Cu$_3$O$_{6+\rm x}$~\cite{fournier1}.

\subsection{Spin splitting}
\begin{figure}[htbp!]
\centering
\includegraphics[width=0.95\textwidth]{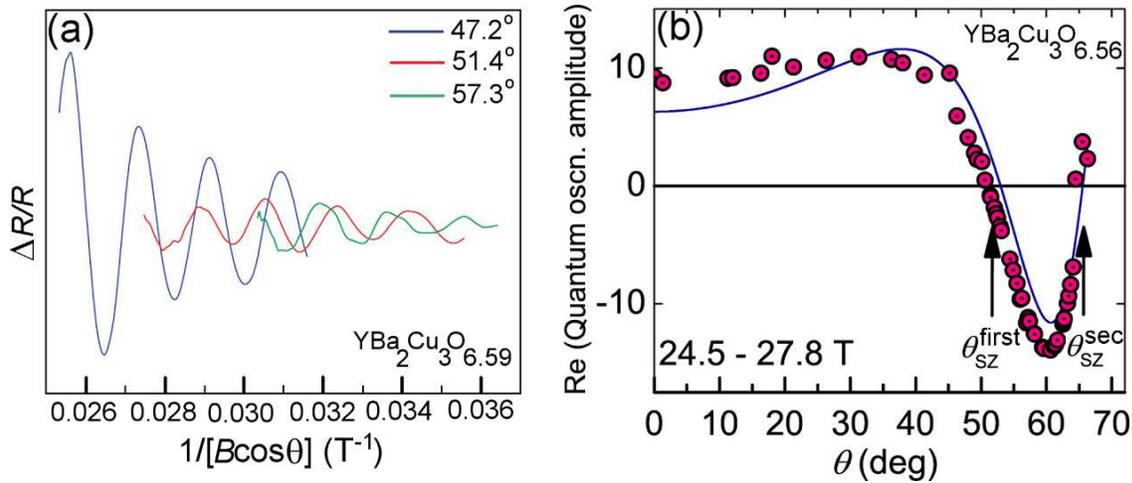}
\caption{(a) Quantum oscillations in the $c$-axis resistance measured in YBa$_2$Cu$_3$O$_{6.59}$ at different angles of inclination of the magnetic field to the $c$-axis, showing the phase flip expected for a spin zero in the vicinity of $50^\circ$, from ref.~\cite{ramshaw1}. (b) Cross correlation of the measured quantum oscillations in the contactless resistivity as a function of the angle of inclination of the magnetic field to the $c$-axis, showing a zero crossing of quantum oscillation amplitude accompanied by a phase inversion at spin zero angles $\theta_{\rm sz}^{\rm first} \approx 53^\circ$ ($r=2$) and $\theta_{\rm sz}^{\rm sec} \approx 64^\circ$ ($r+1=3$)~\cite{sebastian5}. The solid line shows a fit to the angle-dependent data for $g^\ast=2.10$ including a finite value of anisotropy, from ref.~\cite{sebastian5}.
}
\label{spinzero}
\end{figure}

Zeeman splitting of the quasiparticles responsible for quantum oscillations in underdoped YBa$_2$Cu$_3$O$_{\rm {6+x}}$ has been verified by angular dependent quantum oscillation measurements. For a finite Zeeman splitting, the quantum oscillation amplitude is anticipated to vary with angle according to $\cos\big(\frac{\pi}{2}\frac{m^\ast g^\ast}{m_{\rm e}\cos{\theta}}\big)$, where $\frac{m^\ast}{m_{\rm e}}$ is the effective cyclotron mass, $g^\ast$ is an effective $g$-factor, and $\theta$ is the angle of inclination between the applied magnetic field and the crystalline $c$-axis~\cite{shoenberg2}. An experimental verification can be provided by locating the crossing of quantum oscillation amplitude from positive to negative through an angle where the amplitude is exactly zero, known as a `spin zero' angle $\theta_r$, defined by $\frac{m^\ast g^\ast}{m_{\rm e}\cos{\theta_r}}=2r+1$, where $r$ is an integer. Furthermore, in order to determine the value of $\frac{m^\ast g^\ast}{m_{\rm e}}$ from spin zeros alone, the location of at least two successive spin zeros corresponding to $r$ and $r+1$ respectively needs to be identified in order to solve for the two unknowns $\frac{m^\ast g^\ast}{m_{\rm e}}$ and $r$ from the above equation. Angle dependent quantum oscillation measurements up to high angles reveal a minimum in amplitude near 50$^\circ$, first identified in ref.~\cite{ramshaw1} (figure~\ref{spinzero}a). In ref.~\cite{ramshaw1}, it was suggested from a single spin zero together with an overall fit to the form of the quantum oscillation amplitude envelope that $g^\ast \approx 2$. Subsequent measurements have identified a first spin zero at 53$^\circ$, and a second spin zero at 64$^\circ$~\cite{sebastian5} (figure~\ref{spinzero}b). The two identified spin zeros at high angles are found to correspond to values of $r=2$, $r+1=3$, yielding a value of $g^\ast = 2.10(5)$. The absence of an additional low angle spin zero indicates an anisotropy of the $g$-factor associated with the Fermi surface pocket, such that the spin susceptibility is $\approx$~50~\% higher for magnetic field aligned parallel to the $c$-axis than for magnetic field aligned perpendicular to the $c$-axis~\cite{sebastian5}. The agreement of the measured anisotropy in the $g$-factor with the anisotropy reported for the planar Cu sites by nuclear magnetic resonance~\cite{walstedt1} is consistent with Fermi surface pockets originating from the CuO$_2$ planes rather than the CuO chains, for which either no anisotropy or a small anisotropy in the opposite direction is reported~\cite{walstedt1}.

\subsection{Fermi surface transformation}
The first indication from quantum oscillations that a dramatic transformation in the electronic structure has taken place in underdoped YBa$_2$Cu$_3$O$_{6+x}$ is the low value of measured frequency representing a Fermi surface enclosing $\approx 2 \%$ of the Brillouin zone~\cite{doiron1,audouard1,sebastian1}, as opposed to a hole Fermi surface enclosing $> 50 \%$ of the Brillouin zone expected from paramagnetic band structure calculations, and measured in overdoped Tl$_2$Ba$_2$CuO$_{6+\rm x}$~\cite{vignolle1} (shown in figure~\ref{phasediagram}). Furthermore, given the hole character of the large Fermi surface on the overdoped side, it is surprising that quantum oscillations occur in conjunction with negative Hall and Seebeck effects in underdoped YBa$_2$Cu$_3$O$_{6+x}$ and YBa$_2$Cu$_4$O$_8$~\cite{leboeuf1}. If one were to neglect the effects of the vortex (liquid or solid) state in which regime quantum oscillations are observed~\cite{sebastian3}, a negative Hall and Seebeck effect would seem to correspond to electron-like cyclotron orbits. The transformation of an electronic structure comprising a large hole-like Fermi surface into an electronic structure comprising Fermi surface sections of small size yielding a negative Hall and Seebeck effect has been sought to be explained in terms of reconstruction of the Fermi surface by translational symmetry breaking~\cite{millis1, chakravarty1, kivelson1}.


\subsection{Single carrier Fermi surface pocket with multiple frequencies}
\begin{figure}[htbp!]
\centering
\includegraphics[width=0.9\textwidth]{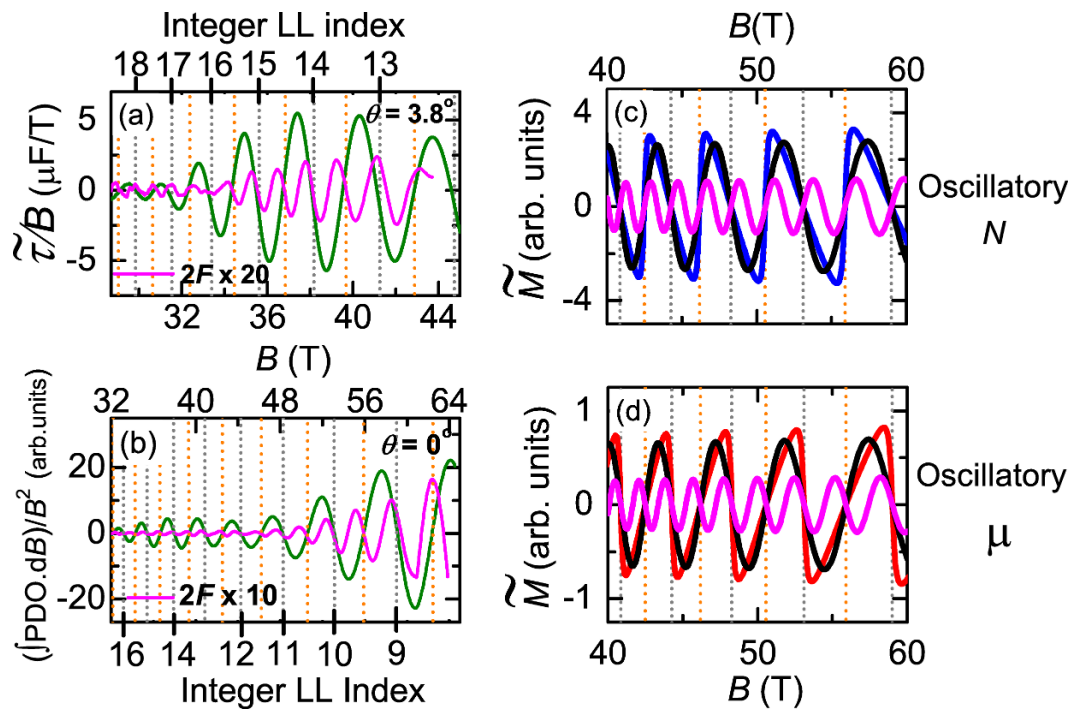}
\caption{(a) Oscillations in the magnetic torque $\tau$, with the extracted second harmonic depicted in purple, from ref.~\cite{sebastian7}. We note that the polarity of the magnetisation curves has been determined experimentally by reference to the field-dependence of the slowly varying diamagnetic background~\cite{sebastian3}, and by comparison with the heat capacity~\cite{riggs1}. Integer Landau level filling factors (at which each Landau level becomes completely filled) are indicated. (b) Measured oscillations in the integrated contactless resistivity (measured by the PDO resonant frequency shift), from ref.~\cite{sebastian7}, which exhibit a similar behavior to those in the magnetic torque with the harmonic indicated again in pink. (c) Simulated inverse sawtooth oscillations in the magnetisation for an oscillatory carrier number $N$ (blue curve) as expected for a multi-section Fermi surface. The fundamental and second harmonic components are shown in black and pink respectively. (d) Simulated sawtooth oscillations in the magnetization for an oscillatory chemical potential $\mu$ (red curve), further separated into fundamental (black curve) and second harmonic (pink curve) components. The phase relation of the experimentally determined second harmonic to the fundamental shown in (a) and (b) is consistent with the oscillatory chemical potential scenario shown in (d).}
\label{chemicalpotential}
\end{figure}

In determining the dramatically transformed electronic structure in the underdoped cuprates, a crucial question pertains to whether the observed quantum oscillations arise from multiple Fermi surface sections in momentum space or from a single carrier Fermi surface pocket with multiple frequencies. The observation of multiple frequencies by quantum oscillations in underdoped YBa$_2$Cu$_3$O$_{6+x}$ ~\cite{sebastian3,audouard1,sebastian4,ramshaw1,sebastian5}, with a spectrally dominant low frequency ($F_\alpha=$~532(5)~T), flanked by two low frequencies (shown in figure~\ref{100T}) and a higher frequency~\cite{sebastian3,sebastian4} could conceivably indicate multiple Fermi surface sections of different carriers and different locations in momentum space~\cite{chakravarty1, sebastian4}. Experiments have only recently distinguished whether these multiple frequencies correspond to multiple sections of Fermi surface~\cite{sebastian7} (see figure~\ref{chemicalpotential}), and whether they are compatible with the existence of both electron-like and hole-like cyclotron orbits. Harmonic analysis of measured quantum oscillations points to a single carrier Fermi surface pocket~\cite{audouard1,ramshaw1,sebastian7} with multiple frequencies likely arising from effects of bilayer splitting, magnetic breakdown and warping~\cite{sebastian7,sebastian8}. This deduction follows from the observation that the behaviour of the harmonic of the dominant oscillations is best understood by a scenario typical of layered metals in which the chemical potential has a strong oscillatory component. Moreover, the strength of the oscillations of the chemical potential needed to understand the harmonic is consistent with the existence of a single Fermi surface section with a density of states at the Fermi level that is dominant over that of any remaining particle reservoir coming from the CuO$_2$ planes, CuO chains or, for example, the electrical leads.

\begin{figure}[htbp!]
\centering
\includegraphics[width=0.8\textwidth]{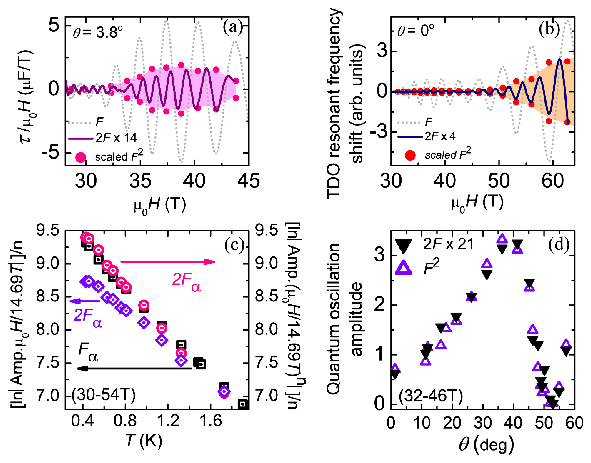}
\caption{Quantum oscillations in the magnetic torque (a) and contactless resistivity (b) revealing the proportionality between the amplitude of the second harmonic (depicted in pink and blue respectively) and the square of the amplitude of the fundamental (red dots) as a function of magnetic field, from ref.~\cite{sebastian7}. The actual fundamental oscillations are shown in light grey in the background. (c) A figure indicating how the amplitude of the harmonic departs from the temperature-dependence anticipated in the Lifshitz-Kosevich theory~\cite{shoenberg2} in which the experimental harmonic mass is twice that of the fundamental. Instead, the amplitude of the harmonic is proportional to the square of that of the fundamental (details provided in reference~\cite{sebastian7}). (d) Comparison of the scaled experimental harmonic amplitude and the square of the fundamental amplitude as a function of angle, from ref.~\cite{sebastian7}, showing excellent agreement. A spin zero which would have been expected at~$\approx$~18$^\circ$ in the second harmonic in the case of oscillations in the carrier density (corresponding to the spin zero observed at $\approx$~53$^\circ$ in the fundamental) is absent, confirming the dominance of oscillations in chemical potential.}
\label{squareproportionality}
\end{figure}

\mbox{}

The first signature of chemical potential oscillations is provided by the relative phase of the fundamental and second harmonic (2$F$) oscillations. Falling regions of the signal for fundamental and second harmonic oscillations measured in magnetisation and the integrated contactless resistivity in underdoped YBa$_2$Cu$_3$O$_{6+x}$ coincide at integer filling factors (see figure~\ref{chemicalpotential}), as characteristic of the sawtooth oscillation form seen in simple two-dimensional metals~\cite{shoenberg1,harrison2, usher1}. The second signature of dominant chemical potential oscillations in YBa$_2$Cu$_3$O$_{6+x}$ is the proportionality of the second harmonic amplitude to the square of the fundamental as a function of angle, magnetic field, and temperature, over the entire region of experimental parameter space in which harmonics can be detected~\cite{sebastian7} (examples provided in figure~\ref{squareproportionality})$-$ harmonics being detected from 30 to 85~T and angles between the crystalline $c$-axis and the magnetic field extending up to 57$^\circ$. Both of these signatures are expected for two-dimensional (or quasi-two-dimensional) metals consisting chiefly of a single Fermi surface section, in which the fixed number of carriers causes the chemical potential to be pinned to the highest occupied Landau level~\cite{shoenberg1,harrison2, usher1} [see figure~\ref{chemicalpotential}(d)]. Furthermore, the observed amplitude envelope of the harmonic, which scales as the square of the fundamental, is inconsistent with opposite carrier pockets which are equivalent in all other aspects~\cite{sebastian4}. The observed quantum oscillations therefore contrast with the expectation for materials with insignificant chemical potential oscillations (typical of multi-section Fermi surfaces, or more generally, of three-dimensional materials~\cite{shoenberg2}.) 

\mbox{}


We are led to the conclusion that quantum oscillations arise chiefly from a single carrier Fermi surface pocket~\cite{sebastian7,notemultiple}. Complementary heat capacity experiments that yield a low value of linear coefficient of specific heat at high magnetic fields are further consistent with the deduction of a single carrier Fermi surface section as opposed to multiple sections of Fermi surface~\cite{riggs1}. Multiple frequency components from the spectrally dominant quasi-two-dimensional pocket likely arise from additional effects such as finite $c$-axis dispersion, bilayer splitting, and magnetic breakdown~\cite{audouard1,ramshaw1,sebastian8}. We note that the existence of additional light chain sections or Fermi surface pockets~\cite{hossain1} that are consistent with the upper bound of the size of chemical potential harmonic oscillations~\cite{sebastian7} may further contribute to the observation of quantum oscillations in the negative Hall effect~\cite{schofield1}.

\begin{figure}[htbp!]
\centering
\includegraphics[width=0.95\textwidth]{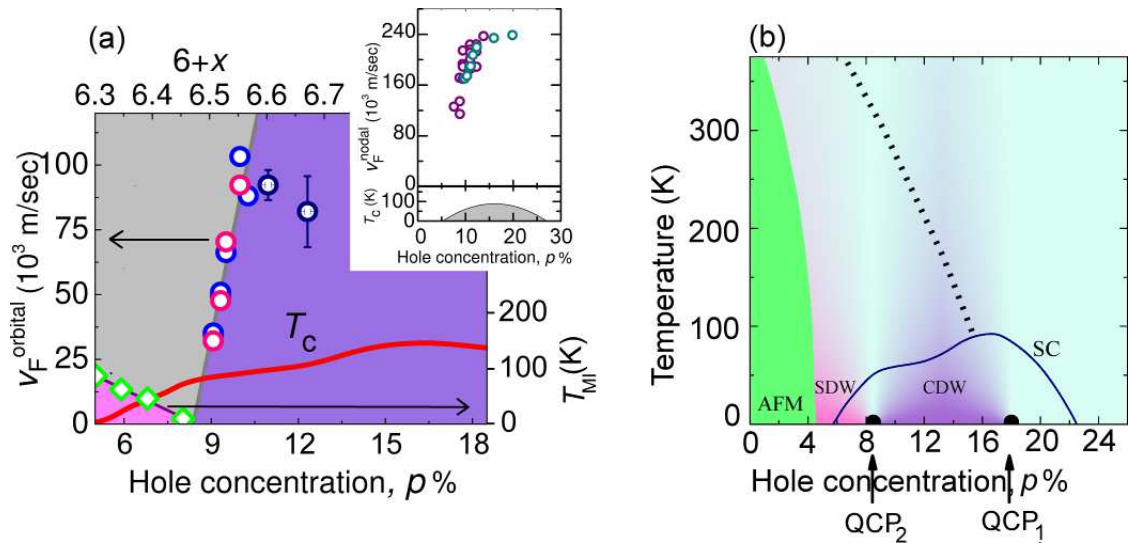}
\caption{
(a) The doping-dependent orbitally-averaged Fermi velocity obtained from quantum oscillation experiments in underdoped YBa$_2$Cu$_3$O$_{6+x}$, from ref.~\cite{sebastian6} (shown as open circles) on the l.h.s axis, estimated using $v_{\rm F}=\frac{\sqrt{2\hbar eF}}{m^\ast}$, where $F$ is the quantum oscillation frequency in tesla and $m^\ast$ is the measured effective mass. The metal-insulator transition temperature ($T_{\rm MI}$) corresponding to an upturn in resistance is shown as open diamonds on the r.h.s. axis~\cite{leboeuf2}. The inset shows the doping-dependent nodal Fermi velocity obtained from high resolution laser ARPES experiments performed on underdoped Bi$_{2}$Sr$_{2}$CaCu$_{2}$O$_{8+\delta}$~\cite{vishik1,anzai1}, estimated from the slope $v_{\rm F}=\frac{\partial E}{\hbar\partial k}$ in the measured dispersion. Cyan and purple circles are obtained from Vishik {\it et al.}~\cite{vishik1} and Anzai {\it et al.}~\cite{anzai1} respectively. (b) Schematic of the phase diagram in YBa$_2$Cu$_3$Cu$_{6+x}$ as a function of hole doping. The superconducting dome at zero magnetic field is denoted by a solid line. Notional regions of spin and charge order (or possibly of short range order) in a magnetic field are denoted in pink and violet respectively ~\cite{haug1,baledent1,julien1}. Neutron scattering and NMR in an applied field both indicate the suppression of spin order at hole dopings higher than $\approx 8.5 \%"$ after refs.~\cite{haug1,baledent1,wu1}. A low doping quantum critical point is denoted as QCP$_2$, potentially demarcating the boundary between spin and charge orders~\cite{haug1,baledent1,julien1,perali1}, while an optimal doping quantum critical point is postulated at QCP$_1$. Multiple orders are likely to be closely competing at each QCP.

}
\label{fermivelocity}
\end{figure}

\subsection{Nodal density of states}

We next turn to the question of the momentum space location of the Fermi surface probed by quantum oscillations, bearing in mind that it consists of a single carrier surface yielding multiple oscillatory components. Experimental evidence obtained using different techniques on the underdoped cuprates points to an electronic density-of-states at the Fermi level concentrated primarily at the nodal location of the Brillouin zone ${\bf k}_{\rm node}=[\pm\frac{\pi}{2a},\pm\frac{\pi}{2b}]$. We review the evidence below. 

\subsubsection{Antinodal pseudogap:}
ARPES experiments enable the momentum-dependent single-electron spectral function at the Fermi energy to be mapped directly~\cite{shen1,hossain1} (shown for the case of underdoped YBa$_2$Cu$_3$O$_{6+x}$ for $x\approx$~0.5 ($p \approx 9 \%$) in the leftmost inset to figure~\ref{phasediagram}a). A pseudogap of size $\approx$~50~meV, also indicated by Knight shift measurements~\cite{alloul1}, opens at the antinodes ${\bf k}_{\rm antinode}=[\pm\frac{\pi}{a},0]$ and $[0,\pm\frac{\pi}{b}]$ in the underdoped cuprates~\cite{norman1}, with the remnant density-of-states at the Fermi level concentrated at `Fermi arcs' located at the nodal locations of the Brillouin zone ${\bf k}_{\rm node}=[\pm\frac{\pi}{2a},\pm\frac{\pi}{2b}]$. Some photoemission experiments report nodal particle-hole asymmetry about the Fermi level~\cite{yang1}. The antinodal pseudogap appears to be ubiquitous in the underdoped cuprates regardless of the specific nature of the ground state, having been observed in systems with reported antiferromagnetism, superconductivity, and stripe ground states in addition to systems where the origin of the pseudogap is unclear~\cite{damascelli1}. Frequency-dependent optical conductivity measurements reveal an absence of a coherent Drude peak along the $c$-axis~\cite{basov1}, also indicating a pseudogap at the antinodes, with little change on introducing a magnetic field as high as 8~T~\cite{tranquada1}.

\begin{figure}[htbp!]
\centering
\includegraphics[width=0.75\textwidth]{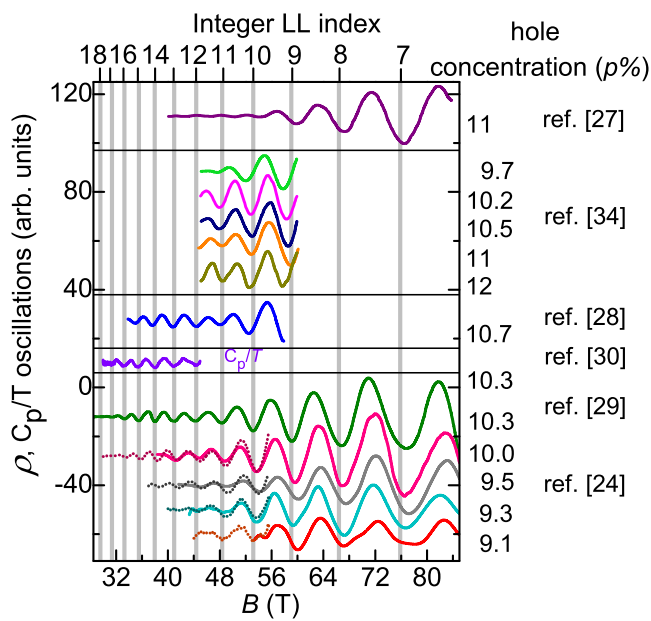}
\caption{Quantum oscillations in underdoped YBa$_2$Cu$_3$Cu$_{6+x}$ over an interval in hole concentration $p \approx 9\% - 12\%$ measured in a range of samples using different transport techniques and heat capacity measurements in multiple magnets by various groups~\cite{sebastian6, singleton1,ramshaw1,sebastian5,riggs1,vignolle2}. The measured oscillations are seen to be very similar in phase, indicating a largely unchanged quantum oscillation frequency across the measured range of hole concentration. The location of integer Landau levels is denoted by light gray vertical lines, and is seen to be similar for all dopings.}
\label{oscillationrange}
\end{figure}

\subsubsection{Collapse in nodal Fermi velocity:}
Crucially relating the observed quantum oscillations to the momentum resolved electronic structure observed by photoemission is the rapid collapse in Fermi velocity observed at low dopings by both techniques~\cite{sebastian1,sebastian6,vishik1,anzai1} (see figure~\ref{fermivelocity}a). The observed quantum oscillations versus field in underdoped YBa$_2$Cu$_3$O$_{6+\rm x}$ remain in phase over a range in dopings between $p=$~9~\% and 12~\%, as measured by various groups using different transport techniques in different high field magnets (shown in figure~\ref{oscillationrange} from refs.~\cite{sebastian6,singleton1,ramshaw1,sebastian5,riggs1,vignolle2}), indicating a largely constant frequency. However, strikingly, measured quantum oscillations observe a steep enhancement in the effective mass of the dominant ($F_\alpha$) frequency by $\approx$ 3 times between dopings $p=$~11~\% and 9~\%~\cite{sebastian6}, extrapolating to a collapse in the orbitally-averaged Fermi velocity at a critical point $p_{\rm c}~\approx$~8.5~\% (or $x\approx$~0.46) in underdoped YBa$_2$Cu$_3$O$_{6+x}$ (see figure~\ref{fermivelocity})~\cite{sebastian1}. Photoemission experiments in Bi$_2$Sr$_2$CaCu$_2$O$_{8+\delta}$ also reveal a precipitous drop in nodal Fermi velocity at a low hole concentration~\cite{vishik1,anzai1}, in qualitative agreement with the critical point yielded by quantum oscillation observations in YBa$_2$Cu$_3$O$_{6+x}$. Indications, therefore, are that ARPES and quantum oscillation measurements both relate to the same region in momentum space (i.e.) the quasiparticles that comprise the Fermi surface pocket observed by quantum oscillations arise from the nodal region at which the concentration of density-of-states at the Fermi level is also observed by ARPES measurements, as opposed to the antinodal region (figures~\ref{phasediagram}a,b). The hole concentration where the Fermi velocity collapses represents a low doping quantum critical point, potentially demarcating the boundary between spin and charge orders~\cite{haug1,baledent1,julien1,perali1}

\subsubsection{Quantum oscillations entering the vortex solid regime:}

\begin{figure}[htbp!]
\centering
\includegraphics[width=0.85\textwidth]{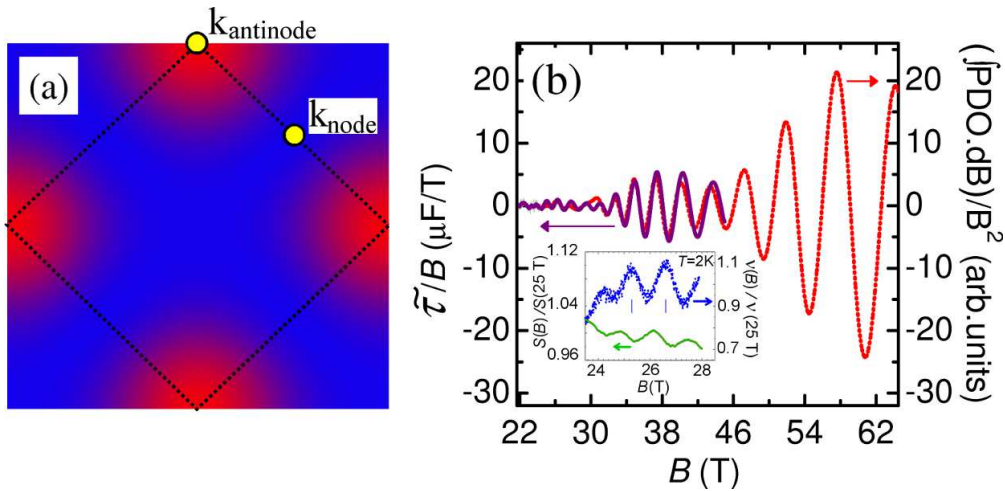}
\caption{(a) A contour plot of the square of the superconducting pairing potential in which blue and red correspond to zero and full squared amplitude respectively. (b) Oscillations measured in the magnetic torque $\tau$ (purple curve, l.h.s. axis) and integrated contactless resistivity measured by a resonant oscillatory shift (PDO) (red curve, r.h.s. axis) in underdoped  YBa$_2$Cu$_3$O$_{6+x}$ ($x=$~0.56) down to low magnetic fields of $\approx 22$~T from ref.~\cite{sebastian7}. The inset shows quantum oscillations in the Seebeck coefficient ($S$) in green on the left axis, and quantum oscillations in the Nernst coefficient ($\nu$) in blue on the right axis measured at $T=2$ K, normalized to their respective values at $B=25$ T extending down to low magnetic fields (from ref.~\cite{laliberte1}).
}
\label{vortex}
\end{figure}

Magnetic quantum oscillations are observed to persist to fields as low as $\approx$~22~T~\cite{laliberte1,sebastian7} in underdoped YBa$_2$Cu$_3$O$_{6+\rm x}$ [see figure~\ref{vortex}(b)], where the d$_{\rm x^2-y^2}$ superconducting gap is well developed [see figure~\ref{vortex}(a)], and is considerably larger than the cyclotron energy. While damping of the cyclotron motion is expected to be small at the nodal region where the superconducting gap is minimum, it is expected to be significantly larger at the antinodal region where the size of the d-wave superconducting gap is maximum~\cite{maniv1,yasui1}, as suggested by prior studies of quantum oscillations in the vortex state of type II superconductors~\cite{corcoran1,harrison1,terashima1}. Antinodal cyclotron orbits would be rendered extremely unlikely by the observation of quantum oscillations down to low magnetic fields in underdoped YBa$_2$Cu$_3$O$_{6+\rm x}$.

\subsubsection{Specific heat from nodal density of states:}

The small value of experimentally measured zero temperature linear heat capacity $\gamma\approx$~4-6~mJmol$^{-1}$K$^{-2}$ up to 45~T~\cite{riggs1} appears challenging to reconcile with a scenario in which nodal pockets are accompanied by antinodal pockets, which would yield a significantly higher value of $\gamma$. Each Fermi surface pocket per layer in the original Brillouin zone is associated with a Sommerfeld coefficient in the normal state of size $\gamma\approx$~5~mJmol$^{-1}$K$^{-2}$. This follows from a contribution to $\gamma$ of the form $\gamma=n_{\rm zone}n_{\rm layer}(\frac{m^\ast}{m_{\rm e}})\times$1.47~mJmol$^{-1}$K$^{-2}$, where $n_{\rm zone}=1$ counts the number of such pockets within the Brillouin zone, $n_{\rm layer}=2$ is the number of copper oxide planes per formula unit, and $\frac{m^\ast}{m_{\rm e}}\approx$~1.6. Nodal accompanied by antinodal pockets would therefore yield a Sommerfeld coefficient of at least $\gamma\approx$~10~mJmol$^{-1}$K$^{-2}$, i.e. more than twice the experimental value of $\gamma$ measured by heat capacity at 45~T. This indication of purely nodal pockets is further supported by the inferred strength of chemical potential oscillations discussed earlier that renders unlikely a significant density-of-states at the Fermi level in addition to that attributed to the spectrally dominant Fermi surface pocket~\cite{sebastian7}.

\begin{figure}[htbp!]
\centering
\includegraphics[width=0.6\textwidth]{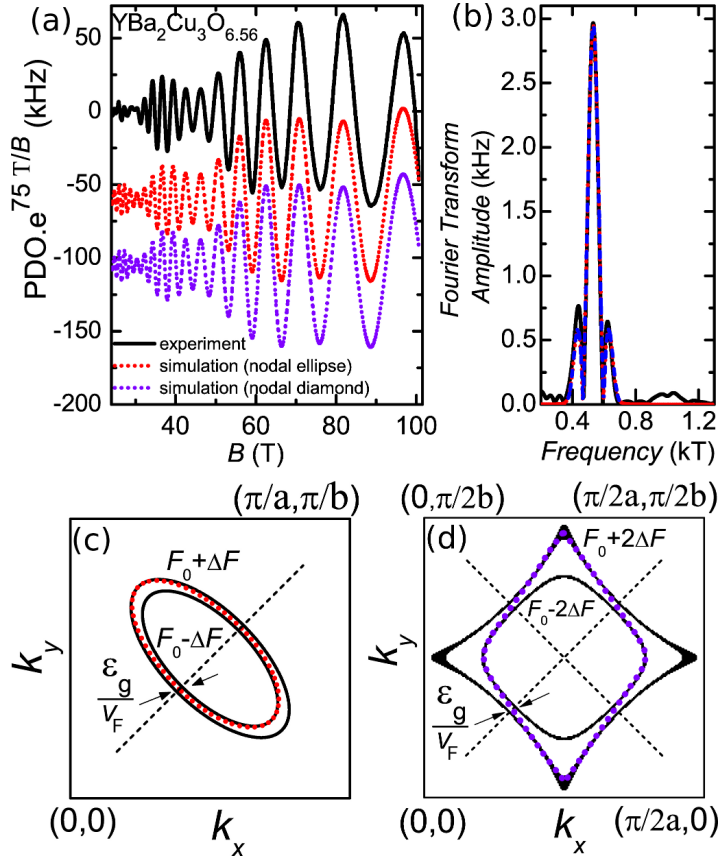}
\caption{(a) Quantum oscillations in the contactless resistivity measured by a resonant shift in frequency (PDO) up to magnetic fields of 101~T (upper solid black line), data from figure~\ref{100T}a~\cite{sebastian8}. Middle and lower dotted lines show fits to bilayer Fermi surface scenarios depicted in (c) and (d) with twofold and fourfold symmetry respectively, allowing for magnetic breakdown tunneling at the nodes. (b) Fourier transform (using
a Blackman window) of the measured quantum oscillation data compared with simulated data from scenarios (c) and (d). The three experimental low frequencies are seen to be reproduced by the simulation with the right ratio of amplitudes.(c) \& (d) Bilayer split Fermi surfaces with twofold and fourfold symmetry with two and four points of magnetic breakdown respectively~\cite{sebastian8}. An example of a magnetic breakdown orbit including tunneling between bilayer-split pockets at the nodes is shown by a dotted line in (c) and (d). $\frac{\varepsilon_{\rm g}}{v_k}$ refers to the size of the magnetic breakdown gap, $F_0$ refers to the central prominent magnetic breakdown frequency of $\approx$~532~T, while $2\Delta F$ (or $4\Delta F$ in the fourfold case) refers to the difference in area between bilayer-split pockets.}
\label{bilayer}
\end{figure}

\subsubsection{Intra-bilayer magnetic breakdown at the nodes:}
Magnetic breakdown occurring at the nodes between Fermi surface sheets from bonding and antibonding bands provides a rather elegant explanation for the beat pattern observed in the quantum oscillation waveform yielding a distinctive Fourier transform consisting of three frequencies with a central prominent peak and two equidistantly spaced side frequencies (fig.~\ref{bilayer})~\cite{sebastian8}. Bilayer coupling would be expected to split a single layer Fermi surface pocket of frequency $F_0$ into two concentric bilayer pockets, as depicted in fig.~\ref{bilayer}(c) and (d), which alone would be insufficient to explain the three-frequency pattern of the Fourier transform. A sufficiently weak bilayer coupling at the nodes, however, would enable magnetic breakdown tunnelling between the split bilayer pockets, yielding a progressive recovery of the single-layer frequency $F_0$ as the strength of the magnetic field is increased~\cite{sebastian8}. Consequently, a three-frequency pattern $F_0-\Delta F$, $F_0$, and $F_0+\Delta F$ is anticipated for sufficiently weak bilayer coupling, which is consistent with the nodal locations at which angular resolved photoemission measurements estimate bilayer coupling $\lessapprox 16$~meV, compared to the estimate of $\approx$150 meV at the antinodal locations of the Brillouin zone~\cite{fournier1}. Fig.~\ref{bilayer}a shows good agreement with experiment for a simulated quantum oscillation waveform corresponding to a nodal pocket (simple cases of either two-fold or four-fold symmetry are shown in figures~\ref{bilayer}(c) and (d) respectively) for a magnetic breakdown field of $\approx$ 10~T~\cite{sebastian8}.

\mbox{}

\section{Electronic structure models and their compatibility with experimental observations}

The initial discovery of quantum oscillations in the underdoped cuprates sparked an abundance of theoretical proposals such as refs.~\cite{lee1,millis1,chakravarty1,kivelson2,varma1,pereg1,wilson1,vafek1,bratkovsky1,yang2,carrington1,chen1,chubukov1,oh1,sushkov1} in an attempt to explain them. The last few years have yielded extensive quantum oscillation data measured by various groups over a wide region of phase space in the underdoped cuprates, serving to constrain feasible theoretical interpretations. Furthermore, complementary experiments that probe the electronic structure in the underdoped cuprates such as photoemission and heat capacity, are crucial to consider in conjunction with the full range of quantum oscillation measurements in examining current theoretical proposals or seeking alternate scenarios to arrive at a comprehensive picture of the electronic structure in the underdoped cuprates. In this section we consider various proposals for the electronic structure in underdoped cuprates, and examine their compatibility both with the above key observations from quantum oscillations, and with complementary experiments.

\subsection{Non-translational symmetry breaking proposals}
Various models have been put forward for the creation of nodal Fermi surface pockets in the normal state of the underdoped cuprates in the absence of translational symmetry-breaking~\cite{varma1,pereg1,wilson1,vafek1,bratkovsky1,yang2}. Such models need to be examined closely in order to discern whether they can yield quantum oscillations with $1/B$ periodicity~\cite{sebastian1}, quasiparticles with Fermi-Dirac statistics~\cite{sebastian2}, angle-dependent spin `zeros' in the quantum oscillation amplitude~\cite{ramshaw1,sebastian5}, as well as negative transport coefficients such as those experimentally observed~\cite{leboeuf1}. 

It has also been suggested that the observed Fermi surface pockets may arise from the non CuO$_2$-planar component of the electronic band structure, such as that from the CuO-chains~\cite{riggs1,carrington1}. Photoemission results, however, show that the band predicted to produce such a chain pocket is located $\approx~0.6$~eV below the Fermi level~\cite{hossain1}. The anisotropy of the $g$-factor associated with the Fermi surface pocket further supports a CuO$_2$ planar origin for the small Fermi surface pockets~\cite{sebastian5}. It is also doubtful whether the density of states at the Fermi level associated with the non-plane component (proposed to be given by the zero-field contribution $\gamma_0 \approx 1.7$ mJmol$^{-1}$K$^{-2}$ to the Sommerfeld coefficient) is sizeable enough to yield the observed amplitude of heat capacity oscillations in the normal state~\cite{riggs1}, given by the expression in ref.~\cite{cp}, and taking into account the size of Dingle damping ($\Gamma \gtrapprox 100$~T in the amplitude prefactor $e^{-\Gamma /B}$)~\cite{ramshaw1,sebastian7}.

\subsection{Translational symmetry breaking proposals}

Alternatively, we take the starting point for the electronic structure of the underdoped cuprates to be a large hole Fermi surface similar to that obtained from band structure calculations and measured in overdoped Tl$_2$Ba$_2$CuO$_{6+\delta}$~\cite{vignolle1}, which may represent an example of a `hidden' Fermi liquid~\cite{anderson2}. Gapping of this large Fermi surface may give rise to the small sections of Fermi surface observed. Potential effects of density wave reconstruction, superconductivity, or a distinct pseudogap mechanism may all be useful to consider in order to understand the quantum oscillations corresponding to small sections of Fermi surface. We start by considering Fermi surface reconstruction by a form of density wave order that breaks translational symmetry to yield small Fermi surface pockets. We consider various forms of order, and examine them for consistency with both quantum oscillations, and complementary experiments.

\begin{figure}[htbp!]
\centering
\includegraphics[width=1\textwidth]{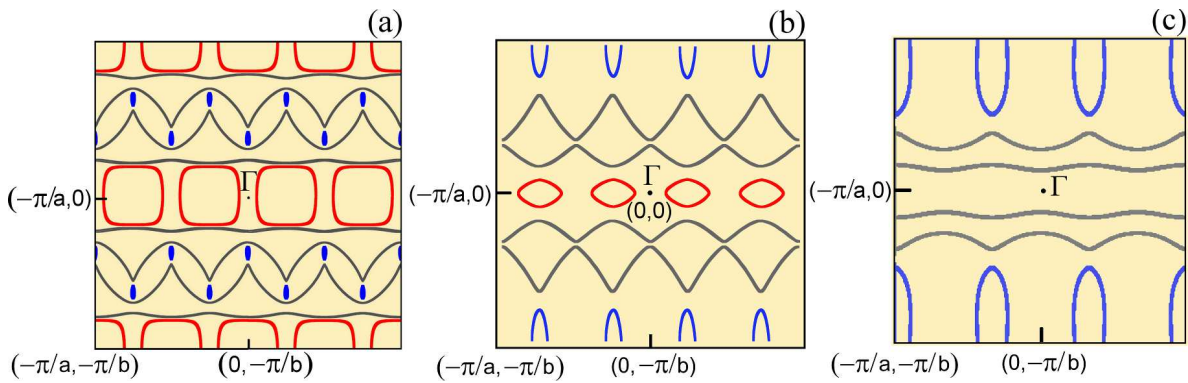}
\caption{Illustrations of possible reconstructed Fermi surfaces for CuO$_2$ planes in underdoped YBa$_2$Cu$_3$O$_{6+\rm x}$. The starting Fermi surface is that shown in the right inset of figure~\ref{phasediagram}a in the conventional Brillouin zone of area $\frac{2\pi}{a} \times \frac{2\pi}{b}$, in YBa$_2$Cu$_3$O$_{6+x}$, $a\approx b\approx$~3.9~\AA. By reduced Brillouin zone, we shall mean below a Brillouin zone with half or less than half this area depending on the size of the ordering wavevector. (a) Example Fermi surface from ref.~\cite{millis1} for an incommensurate (spin-charge) stripe model where the spin potential based on ordering wavevectors of the form ${\bf Q}=[(1\pm2\delta)\frac{\pi}{a},\frac{\pi}{b}]$ is collinear with $\delta=\frac{1}{8}$ (for the example shown here, the accompanying charge potential with charge ordering wavevector twice that of the spin ordering wavevector is 0) yielding one electron pocket near the antinodal location (shown in red) potentially accompanied by two small hole pockets at the nodes (shown in blue), together with open sheets at the nodal locations (shown in black). (b) Example Fermi surface from ref.~\cite{kivelson1} for a charge density wave (i.e. a charge stripe) with ordering wavevectors of the form ${\bf Q}_{x}=[\frac{2\pi}{a\lambda_a},0]$ in which $\lambda_a=4$ accompanied by an orthorhombic lattice distortion or electron nematicity. A single electron pocket per reduced Brillouin zone is created at the antinodal location (shown in red), potentially accompanied by a single hole pocket per reduced Brillouin zone near the other antinodal location (shown in blue), together with open sheets at the nodal locations (shown in gray). (c) Example Fermi surface for a charge-density wave with ordering wavevectors of the form ${\bf Q}_{x}=[\frac{2\pi}{a\lambda_a},0]$ in which $\lambda_a=3$ gives rise to a hole pocket near the antinodal location together with open sheets at the nodal locations. The reconstructed Fermi surface is shown within the full original Brillouin zone in all cases. 
}

\label{afmsmall}
\end{figure}

\begin{figure}[htbp!]
\centering
\includegraphics[width=1\textwidth]{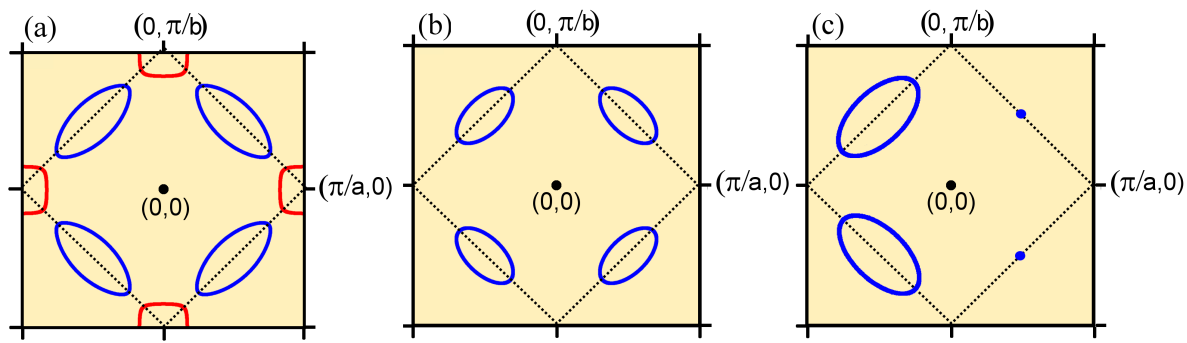}
\caption{
(a) Schematic showing how translation of the Fermi surface by ordering wavevectors of the type ${\bf Q}=[\frac{\pi}{a},\frac{\pi}{b}]$ leads to the creation of hole pockets (blue) situated at ${\bf k}_{\rm nodal}=[\pm\frac{\pi}{2a},\pm\frac{\pi}{2b}]$ and electron pockets (red) at ${\bf k}_{\rm antinodal}=[\pm\frac{\pi}{a},0]$ and $[0,\pm\frac{\pi}{b}]$ when the coupling is sufficiently weak. (b) Schematic showing how translation of the Fermi surface by ${\bf Q}=[\frac{\pi}{a},\frac{\pi}{b}]$ leads to solely hole pockets (blue) at ${\bf k}_{\rm nodal}=[\pm\frac{\pi}{2a},\pm\frac{\pi}{2b}]$ when the coupling is sufficiently large~\cite{lee1}. (c) Schematic showing how translation of the Fermi surface in a spiral state ${\bf Q}=[(1-2\delta)\frac{\pi}{a},\frac{\pi}{b}]$ leads to an asymmetry in the size of nodal hole pockets located at ${\bf k}_{\rm nodal}=[\pm\frac{\pi}{2a},\pm\frac{\pi}{2b}]$ ~\cite{lee1,sushkov1}, with a sufficiently large value of $\delta$ causing half the pockets to disappear, leaving behind half of the nodal hole pockets at $[-\frac{\pi}{2a},\pm\frac{\pi}{2b}]$ (blue)~\cite{sushkov1}.}
\label{afmbig}
\end{figure}

\subsubsection{Electronic structure proposals $-$ quantum oscillations from an antinodal Fermi surface section:}

We first consider electronic structure proposals postulating chiefly antinodal Fermi surface sections, in which an antinodal electron pocket is proposed to give rise to quantum oscillations measured in the underdoped cuprates.

\paragraph{Large $\bf Q$ order:}
In the case of collinear antiferromagnetic order with spin ordering wavevector of the type ${\bf Q}=[(1\pm2\delta)\frac{\pi}{a},\frac{\pi}{b}]$ (for instance with $\delta \approx 1/8$ as reported for other cuprate families~\cite{tranquada2}), potentially accompanied by a charge potential (i.e. spin-charge stripes), open Fermi surface sheets are produced at the nodal region, accompanied by electron pockets at the antinodal region of the Brillouin zone [see figure~\ref{afmsmall}(a)]~\cite{millis1}. The electron pockets at the antinodal locations have been proposed to correspond to the Fermi surface observed by quantum oscillation measurements, also yielding negative Hall and Seebeck coefficients as experimentally measured~\cite{leboeuf1}.

\paragraph{Small $\bf Q$ order:}
Purely one-dimensional small $\bf Q$ (charge stripe) order with ${\bf Q}_{x} = [\frac{2\pi}{a\lambda_a},0]$ ($\lambda_a\approx4$) would give rise to open Fermi surface sheets~\cite{millis1,harrison3}, while the additional involvement of moderately strong C4 symmetry breaking by lattice orthorhombicity or an electronic smectic phase would yield a small electron pocket located at the antinodal region (${\bf k}_{\rm antinode}=[\pm\frac{\pi}{a},0]$), possibly accompanied by a small hole pocket at the antinodal location (${\bf k}_{\rm antinode}=[0,\pm\frac{\pi}{b}]$)  region of the Brillouin zone in addition to the open Fermi surface sheets extending through the nodal region (see figure~\ref{afmsmall}b) ~\cite{kivelson1}. In this proposal, the electron pockets at the antinodal locations have been suggested to correspond to the Fermi surface observed by quantum oscillation measurements, also yielding negative Hall and Seebeck coefficients as experimentally measured~\cite{leboeuf1}. A purely one-dimensional charge stripe with a smaller value of $\lambda_a$ on the other hand, would yield a hole pocket near the antinodal location accompanying open Fermi surface sheets at the nodal location of the Brillouin zone (the example of $\lambda_a \approx 3$ is shown in figure~\ref{afmsmall}c).

\mbox{}

Given the large antinodal pseudogap observed by photoemission experiments in zero magnetic field~\cite{shen1,hossain1,basov1,tranquada1}, a significant reconfiguration of the electronic structure by the magnetic field ($B\gtrapprox$~22~T) would seem to be required for Fermi surface models involving antinodal pockets to be responsible for the observed quantum oscillations. Even were we to consider that such a magnetic field-induced reconfiguration could result in a significant antinodal density of states at the Fermi level in underdoped YBa$_2$Cu$_3$O$_{\rm 6+x}$~\cite{micklitz1,senthil1}, further problems would remain at high magnetic fields due to the fact that all models proposed involving antinodal Fermi surface pockets would also involve a nodal density of states at the Fermi level. Consequently, such models would be difficult to reconcile with the harmonic enrichment already discussed that supports a dominant contribution to the density of states at the Fermi level from a single closed Fermi surface section (albeit supporting multiple frequency components~\cite{sebastian8,audouard1}). Furthermore an antinodal electron pocket giving rise to the observed quantum oscillations would alone yield a contribution to the Sommerfeld ratio $\gamma$ of 5~mJmol$^{-1}$K$^{-2}$, the additional involvement of nodal open or closed orbits also with a finite contribution to the linear coefficient of heat capacity would be difficult to reconcile with $\gamma\approx$~4-6~mJmol$^{-1}$K$^{-2}$ observed in bulk heat capacity studies in the same field, temperature and doping ranges of the quantum oscillation experiments~\cite{riggs1} (the lower value of 4~mJmol$^{-1}$K$^{-2}$ is measured relative to the zero field limit of the Sommerfeld coefficient).


\subsubsection{Electronic structure proposals: quantum oscillations from a nodal Fermi surface section:}
Electronic structure models in which quantum oscillations arise from Fermi surface pockets situated at the nodal locations of the Brillouin zone have been proposed to provide an explanation for the majority of experimental observations relating to the electronic structure in the underdoped cuprates.

\paragraph{Large $\bf Q$ order:}
Long range order with ordering wavevectors of the type $[\frac{\pi}{a},\frac{\pi}{b}]$, corresponding to translational symmetry breaking orders such as a d-density wave~\cite{chakravarty1} or antiferromagnetism~\cite{chen1,chubukov1,oh1} would result in Fermi surface reconstruction yielding hole pockets at the nodal (${\bf k}_{\rm node}=[\pm\frac{\pi}{2a},\pm\frac{\pi}{2b}]$) regions of the Brillouin zone. Commensurate order in which ${\bf Q}=[\frac{\pi}{a},\frac{\pi}{b}]$ yields two hole pockets in the reduced Brillouin zone per CuO$_2$ plane (schematic shown in figure~\ref{afmbig}(b)), yielding a minimum contribution to the Sommerfeld coefficient of ~$\approx$~10~mJmol$^{-1}$K$^{-2}$, which is larger than the $\approx$~4-6~mJmol$^{-1}$K$^{-2}$ observed in bulk heat capacity studies~\cite{riggs1}.

\begin{figure}[htbp!]
\centering
\includegraphics[width=0.75\textwidth]{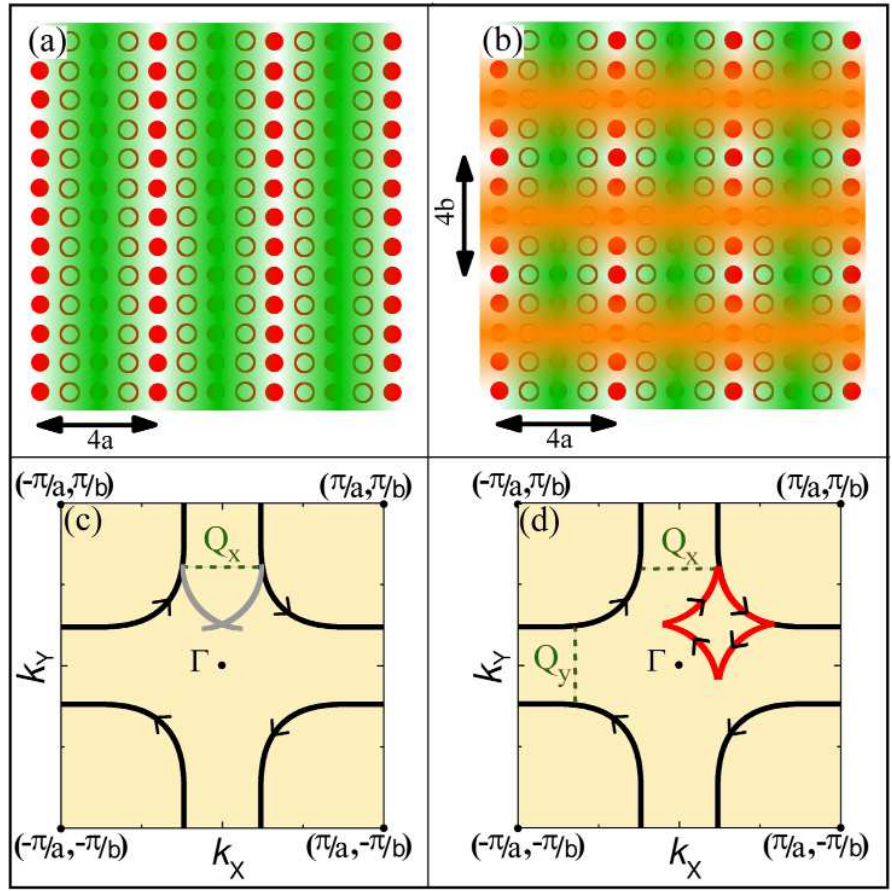}
\caption{(a) Schematic of period $4a$ charge ordering (along the $x$ direction), where $a$ is the lattice constant. (b) Schematic of period $4a$ and $4b$ charge orderings (along $x$ and $y$) where $a$ and $b$ are lattice constants shown in green and orange respectively. Different colours are used to represent the possibility that order occurring along the $b$ axis may have a different amplitude or phase relative to that along $a$ axis. (c) Schematic showing the effect of a unidirectional charge ordering vector of the type ${\bf Q}_x=[\frac{\pi}{2a},0]$ spanning the flat parts of the Fermi surface, which would lead to open Fermi surface sheets at the nodal locations~\cite{millis1,harrison3}. (d) Schematic showing how biaxial charge order would lead to nodal Fermi surface pockets (shown by red line), potentially yielding `Fermi arcs' located in the nodal region~\cite{harrison3}. We will take as convention anticlockwise cyclotron motion for orbits enclosing unfilled (hole) states. The nodal Fermi surface pocket from biaxial charge order has clockwise cyclotron rotation, opposite to the original closed orbits shown in black.}
\label{chargeschematic}
\end{figure}

For spiral order with finite incommensurability parameter $\delta$, nodal hole pockets that are asymmetric in size are produced, with half the pockets disappearing for a large enough value of $\delta$ (schematic shown in fig.~\ref{afmbig}(c))~\cite{sebastian4}. While this scenario is intriguing to consider, we note the caveat that there is thus far no evidence for ${\bf Q}=[(1\pm2\delta)\frac{\pi}{a},\frac{\pi}{b}]$ forms of long range spin order in sample compositions (i.e. YBa$_2$Cu$_3$O$_{6+x}$ where 0.69~$>x>$~0.46, equivalent to a hole concentration 12$\%>p>9\%$) in which quantum oscillations are observed~\cite{haug1,julien1}. Neutron scattering experiments reveal elastic and/or inelastic features associated with spin degrees of freedom with ${\bf Q}$ of the type $[(1\pm2\delta)\frac{\pi}{a},\frac{\pi}{b}]$ for $p\lessapprox$~8.5~\% in YBa$_2$Cu$_3$Cu$_{6+x}$~\cite{haug1,baledent1}, which suggest that the region of doping where incipient form of such order occurs ends at the doping above which quantum oscillations begin to be observed~\cite{haug1}, potentially separated by a QCP (schematic phase diagram shown in figure~\ref{fermivelocity}b, similar theoretical suggestions have been made in ref.~\cite{perali1}). Furthermore, given that the resulting nodal pockets are hole-like in character, with a direction of cyclotron orbit rotation that is the same as the direction of cyclotron orbit rotation of the original unreconstructed large hole Fermi surface (figure~\ref{chargeschematic}d), it is unclear as to whether this scenario is compatible with the observation of a negative Hall and Seebeck coefficient in strong Landau-quantizing magnetic fields~\cite{leboeuf1,laliberte1}. While for weak enough density wave amplitude, these models would also yield electron pockets at the antinodal (${\bf k}_{\rm antinode}=[\pm\frac{\pi}{a},0]$ and $[0,\pm\frac{\pi}{b}]$) regions of the Brillouin zone [see figure~\ref{afmbig}(a) and (b)], these antinodal pockets would be gapped away for a sufficiently large density wave amplitude (see figure~\ref{afmbig}(c)). 

\begin{figure}[htbp!]
\centering
\includegraphics[width=0.75\textwidth]{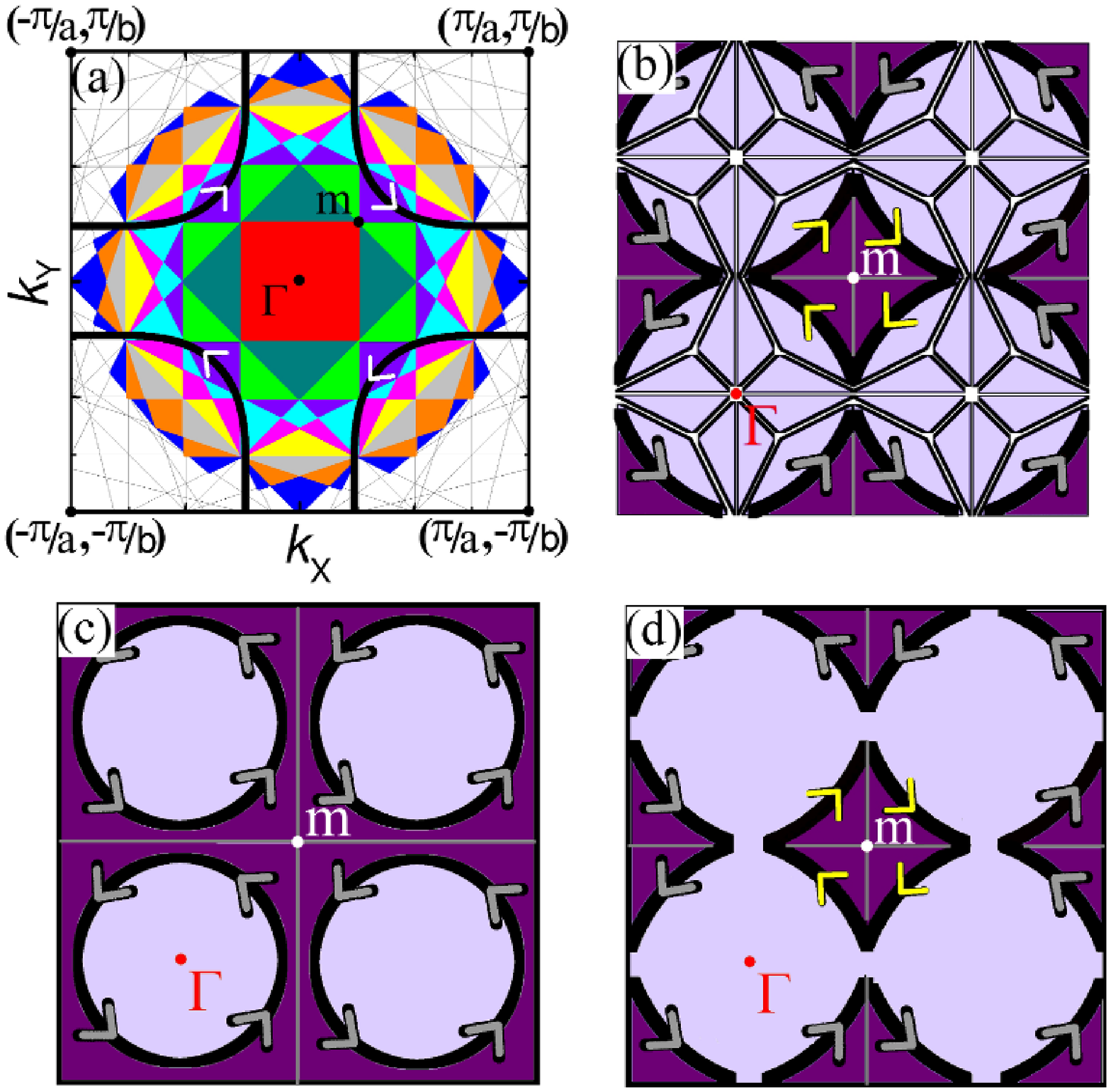}
\caption{(a) A schematic showing the effect of introducing a new $\lambda_a\times\lambda_b$ superstructure on the Fermi surface of underdoped YBa$_2$Cu$_3$O$_{6+x}$, for example, as seen in resonant x-ray scattering experiments~\cite{letacon1}. For simplicity, the case where $\lambda_a=\lambda_b=$~4 is considered. The first ten Brillouin zones of the superlattice are shown. While the flatter portions of Fermi surface straddle multiple Brillouin zone boundaries, the 4 `arcs' reside within the 4$^{\rm th}$ Brillouin zone (purple sections), suggesting that they will be relatively weakly affected by the opening of additional band gaps. Similar to figure~\ref{chargeschematic}, we will take as convention anticlockwise cyclotron motion for orbits enclosing unfilled (hole) states. (b) Schematic showing how the reconstructed Fermi surface is obtained by reassembling the pieces of the 4$^{\rm th}$ Brillouin zone into the 1$^{\rm st}$ Brillouin zone of the superlattice (identified as the reduced Brillouin zone), giving rise to a Fermi surface pocket located at the $m$ point. Filled states are shown in dark purple, and unfilled states are shown in pale violet, a section of the Brillouin zone is enlarged for clarity. More detailed band calculations are presented in ref.~\cite{harrison3}. (c) Schematic showing how a low hole doping ($p\lessapprox$~8~\%) would lead to a hole pocket (i.e. comprising unfilled states) shown in pale violet (i.e. having anticlockwise rotation analogous to the original unreconstructed closed orbits shown in black in (a)) centred at the $\Gamma$ point in the Brillouin zone. (d) Schematic showing how an increased hole doping (relevant to quantum oscillation studies in which $p\gtrapprox$~8~\%) causes the Fermi surface pockets shown in (c) to extend beyond the boundary of the reduced Brillouin zone, giving rise instead to a Fermi surface pocket with electron-like character (i.e. comprising filled states) shown in dark violet, having clockwise rotation opposite to the original closed orbits shown in (a) and centered at the $m$ point in the Brillouin zone.}
\label{brillouinzone}
\end{figure}

\paragraph{Small ${\bf Q}$ order:}
Another possibility for translational symmetry breaking order yielding nodal pockets can also be considered $-$ small $\bf Q$ charge order (or in some cases slowly fluctuating order) appears to be ubiquitous in the underdoped cuprates. Charge ordering at small wavevectors of the type ${\bf Q}_x\approx[\frac{2\pi}{a\lambda_a},0]$ and/or ${\bf Q}_y\approx[0,\frac{2\pi}{b\lambda_b}]$ (where $\lambda_a=\lambda_b\approx~3 - 6$ is the period of the modulation in lattice constants), has been identified in various experiments performed on cuprates, including scanning tunneling microscopy (STM)~\cite{mcelroy1,hoffman1,hanaguri1}, neutron scattering~\cite{tranquada1}, x-ray diffraction~\cite{liu1}, nuclear magnetic resonance~\cite{julien1}, and resonant x-ray scattering (RXS)~\cite{letacon1}.  Recent RXS experiments performed on underdoped YBa$_2$Cu$_3$O$_{6+\rm x}$ in zero magnetic field reveal charge ordering wavevectors $\bf Q_x$ and $\bf Q_y$ along both $x$ and $y$ directions, with a periodicity corresponding to $\lambda_a=\lambda_b\approx$~3.2 ~\cite{letacon1}. Nuclear magnetic resonance (NMR) experiments performed on underdoped YBa$_2$Cu$_3$Cu$_{6+x}$ in 28~T reveal a line splitting consistent with charge order, with a proposed ${\bf Q}_x$ superstructure with $\lambda_a=$~4~\cite{julien1} on assuming an undistorted sinusoidal modulation. X-ray diffraction experiments on optimally-doped YBa$_2$Cu$_3$Cu$_{6+x}$ containing ortho-IV domains of nominal composition $p\approx$~14~\% (i.e. $x\approx$~0.75) indicate signatures of a ${\bf Q}_y$ superstructure with $\lambda_b\approx$~4.7~\cite{liu1} $-$ potentially supported by STM~\cite{edwards1,maki1}, NMR~\cite{kramer1,grevin1} and neutron scattering measurements of phonon softening ~\cite{reznik1,mook2} which persist deep into the underdoped regime~\cite{mook1}.

\begin{figure}[htbp!]
\centering
\includegraphics[width=0.9\textwidth]{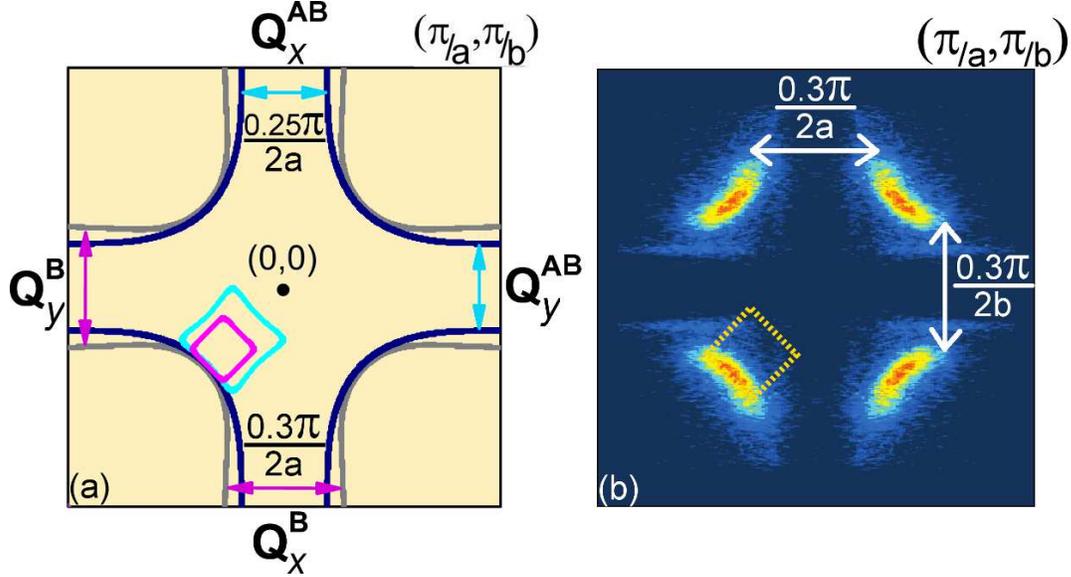}
\caption{(a) Schematic Fermi surface of YBa$_2$Cu$_3$O$_{6+x}$ from a band structure (similar to that in ref.~\cite{elfimov1}) including bilayer splitting into bonding (gray) and antibonding (black) bands for hole concentration $p \approx 9 \%$. The connecting wavevectors, $\bf Q_x^{\rm AB} = [\frac{2\pi}{a \lambda_a^{\rm AB}},0]$, $\bf Q_x^{\rm B} = [\frac{2\pi}{a \lambda_a^{\rm B}},0]$, $\bf Q_y^{\rm AB} = [0,\frac{2\pi}{b \lambda_b^{\rm AB}}]$, $\bf Q_y^{\rm B} = [0,\frac{2\pi}{b \lambda_b^{\rm B}}]$ are shown for the antibonding (AB) and bonding (B) bands respectively. As seen from the figure, $\lambda^{\rm AB} \approx 4$ for the antibonding band is larger than $\lambda^{\rm B} \approx 3.3$ for the bonding band (the precise values of these $\bf Q$ connecting wavevectors depend on specific details of the unreconstructed band structure). The smaller $\lambda^{\rm B}$ coincides with the recent RXS results~\cite{letacon1}. The resulting `electron-like' nodal pocket from biaxial translation of the antibonding bands by $\bf Q_x^{\rm AB}$ and $\bf Q_y^{\rm AB}$ is shown in pink, and from the biaxial translation of the bonding bands by  $\bf Q_x^{\rm B}$ and $\bf Q_y^{\rm B}$ is shown in blue~\cite{harrison5}. The area of the nodal Fermi surface pocket thus yielded, modified by the effect of bilayer splitting and magnetic breakdown~\cite{sebastian8}, is consistent with the measured quantum oscillation frequencies with a spectrally dominant frequency $F=\frac{\hbar}{2\pi e}A_k\approx$~500~T and two equidistantly spaced side frequencies~\cite{bondingnote}. Note that in figure~\ref{nesting}, the difference between antibonding and bonding wavevectors is not shown. (b) `Fermi arcs' as seen in underdoped YBa$_2$Cu$_3$O$_{6+x}$ (the same composition range in which quantum oscillations are observed) in which the surface is treated with potassium, from ref.~\cite{hossain1}. Upon translation of the `Fermi arcs' by ${\bf Q}_x=[\frac{2\pi}{a\lambda_a},0]$ and ${\bf Q}_y=[0,\frac{2\pi}{b\lambda_b}]$ (where $\lambda_a=\lambda_b\approx$~3.2), a nodal `electron-like' pocket is created with area corresponding to a quantum oscillation frequency $F=\frac{\hbar}{2\pi e}A_k\approx$~500~T similar to that measured by quantum oscillation experiments. Portions of figure reproduced with permission from~\cite{hossain1}. 
}
\label{nesting}
\end{figure}

A purely one-dimensional charge order ${\bf Q}_x=[\frac{2\pi}{a\lambda_a},0]$ (see figure~\ref{chargeschematic}(a)) would give rise to open Fermi surface sheets at the nodal locations~\cite{millis1,harrison3} (e.g. of $\lambda_a=4$ shown in figure~\ref{chargeschematic}(c)). However, a superstructure involving simultaneous order along both the $x$ and $y$ directions (see schematic in figure~\ref{chargeschematic}(b)) would lead to a significant reduction in the area of the reduced Brillouin zone area, yielding a nodal Fermi surface pocket comprising filled (as opposed to unfilled or hole) states in the new Brillouin zone~\cite{harrison3,li1} [see figure~\ref{chargeschematic}(d)], potentially accompanied by other pockets depending on the size of the charge density wave amplitude. Fig.~\ref{chargeschematic} shows the simple example in which $\lambda_a = \lambda_b = 4$, where the two perpendicular low-$\bf Q$ wavevectors result in a cyclotron path that traverses a small nodal Fermi surface made up of four nodal regions from the original large Fermi surface as a consequence of Bragg reflection. This construction would yield similar results for different values of the $\bf Q$ wavevector. For instance, the RXS wavevector for $\lambda_a=\lambda_b=3.2$ would connect the tips of the Fermi arcs, or the flat edges of the bonding bands of the large unreconstructed paramagnetic Fermi surface~\cite{letacon1,bondingnote}, also yielding a small nodal Fermi surface section comprising filled states (shown in fig.~\ref{nesting}). In the latter scenario, a different connecting wavevector would correspond to the antibonding bands (see Fig.~\ref{nesting}b), suggesting the possibility of a secondary charge modulation. Different wavevectors for the bonding and antibonding bands could potentially explain the range of $\lambda$ values seen in YBa$_2$Cu$_3$O$_{6+{\rm x}}$ using different experiments~\cite{liu1, letacon1, julien1,edwards1,maki1,reznik1,mook2,mook1}.

\section{Properties of nodal pocket from bilayer low-$\bf Q$ charge order}

The nodal Fermi surface pocket yielded by biaxial low-$\bf Q$ Fermi surface reconstruction has several appealing features consistent with experiment. For this reason we give it special attention.

\subsection {Momentum space location of Fermi surface pocket}
The nodal location of the Fermi surface pocket created by biaxial low-$\bf Q$ order is consistent with (i) the nodal density-of-states at the Fermi level observed as `Fermi arcs' by photoemission experiments in zero magnetic field and the antinodal pseudogap observed by photoemission~\cite{norman1}, optical conductivity~\cite{basov1}, nuclear magnetic resonance knight shift~\cite{alloul1}, and other measurements at zero magnetic field, and (ii) the large superconducting gap at the antinodal region compared to the small size of the superconducting gap at the nodal region of the Brillouin zone.

\subsection {Sommerfeld coefficient}
A solely nodal Fermi surface pocket, unaccompanied by a significant antinodal density-of-states at the Fermi level would result from a large size of density wave matrix element in a biaxial low-$\bf Q$ order scenario, or an additional pseudogap mechanism that gaps the antinodal density of states $-$ as discussed in section 3.2.2.2. A solely nodal pocket created for ${\bf Q}_x=[\frac{2\pi}{a \lambda_a},0]$ and ${\bf Q}_y=[0,\frac{2\pi}{b \lambda_b}]$ would yield a Sommerfeld coefficient in the normal state of size $\gamma\approx$~5~mJmol$^{-1}$K$^{-2}$, in agreement with the small value of measured linear coefficient of heat capacity at high magnetic fields.

\subsection {Sign of the Hall coefficient and Seebeck effect}
The cyclotron orbital motion of the nodal Fermi surface pocket yielded by biaxial charge order is seen to be in the opposite direction to the original unreconstructed hole orbit (figure~\ref{brillouinzone},\ref{chargeschematic}). This is a consequence of the nodal Fermi surface pocket yielded by biaxial low $\bf Q$ charge order comprising unfilled regions of the Brillouin zone as shown in figure~\ref{brillouinzone}. The Hall and Seebeck coefficients may therefore be anticipated to be negative in the high field limit, in agreement with the experimental observation of quantum oscillations in the negative Hall coefficient and Seebeck effect (not considering effects of any background vortex contribution)~\cite{leboeuf1,laliberte1}.

\subsection {Observation of spin splitting}
Planar Fermi surface reconstruction by a biaxial low-$\bf Q$ charge density wave is anticipated to retain a significant Zeeman splitting, as opposed to Fermi surface reconstruction involving certain forms of spin density wave where the Zeeman splitting is expected to be suppressed due to the locking of spin orientation by the spin density wave~\cite{revaz1}. The experimental observation of multiple spin zero angles in the angular dependence of the measured quantum oscillations~\cite{ramshaw1,sebastian7} corresponding to a sizeable Zeeman splitting, and a large value of anisotropic effective `g-factor' $g^\ast$ is therefore consistent with planar Fermi surface reconstruction by a low-$\bf Q$ charge density wave.

\subsection {Luttinger's theorem}

Next we consider the relationship expected by Luttinger’s theorem~\cite{luttinger1} between the area $A_k$ of the Fermi surface pocket in the reduced Brillouin zone of area $A_{\rm {RBZ}}=A_{\rm {BZ}}/(\lambda_{\rm a} \times \lambda_{\rm b})$, and the area enclosing filled states, $\nicefrac{1}{2}(1 - p)A_{\rm BZ}$, in the original Brillouin zone of area A$_{\rm {BZ}}$. For the number of completely filled bands ($n$) given by the floor of $\nicefrac{1}{2}(1 - p)(\lambda_a\times\lambda_b)$, we expect the area enclosing occupied states in the original Brillouin zone to be related to $A_k$ (where $0 < A_k < A_{\rm RBZ}$), by
\begin{eqnarray}
\nicefrac{1}{2}(1 - p)A_{\rm BZ} = n A_{\rm RBZ} + A_k
\label{luttinger}
\end{eqnarray}

\noindent Here $p$ is the hole doping fraction ($0 < p < 1$), and $A_{\rm BZ}$ and $A_{\rm RBZ}$ are the areas of the original and reduced Brillouin zones, respectively. 

Recalling that $A_k$ is the area of filled states, for $A_k \gtrapprox 50 \%$ the cyclotron orbit would be closed around the $\Gamma$ point of the Brillouin zone (as shown in figure~\ref{brillouinzone}(c)), with cyclotron rotation in the same direction to that of the large closed Fermi surface in the original Brillouin zone, and yielding a quantum oscillation frequency $F=(\frac{\hbar}{2\pi e})(A_{\rm {RBZ}} - A_k$). For  $A_k \lessapprox 50 \%$ the cyclotron orbit would be closed around the $m$ point of the Brillouin zone (as shown in figure~\ref{brillouinzone}(d)), with cyclotron rotation in the opposite direction to that of the large closed Fermi surface in the original Brillouin zone, and yielding a quantum oscillation frequency $F=(\frac{\hbar}{2\pi e})A_k$. Thus, for example, if $\lambda_a = \lambda_b = 4$, then for $n$ = 7, $A_k = (1 - 8p) A_{\rm RBZ}$, yielding a quantum oscillation frequency $F=\frac{\hbar}{2\pi e}A_k\approx$~500~T  for $p \approx 9 \%$. For this value of $p$, $A_k$ is equal to $0.28 A_{\rm RBZ}$ and hence it encloses filled states in less than half of reduced Brillouin zone and the cyclotron rotation is expected to be in the opposite direction to that of the large unreconstructed closed Fermi in the original Brillouin zone.

The quantum oscillation frequency in this model is sensitive to both hole concentration $p$ and the value of $\lambda$. To account for the weak variation of quantum oscillations frequency with hole concentration (figure~\ref{oscillationrange}), despite a significant $p$ dependence of the cyclotron effective mass (figure 7), an increase in $\lambda$ with increasing $p$ is indicated~\cite{harrison4}. Evidence for a rise in $\lambda$ with $p$ has indeed been seen via STM measurements~\cite{wise1}, however there is not yet a quantitative demonstration that the $p$ dependence of $\lambda$ is fully consistent with the observed weak $p$ dependence of the area of the observed Fermi surface pocket.

\section{Summary and Outlook}

In summary, quantum oscillation measurements observe a single carrier small Fermi surface section yielding multiple oscillatory components in underdoped YBa$_2$Cu$_3$O$_{6+x}$ at high magnetic fields, which is drastically reduced in size from the overdoped side. A synthesis of complementary experimental techniques and comparison with alternative theoretical models narrows viable electronic structure proposals in the underdoped cuprates to those in which quantum oscillations arise from a nodal Fermi surface pocket. This conclusion is supported in particular by (i) the dramatic reduction in size of Fermi surface observed by quantum oscillation measurements in going from the overdoped to underdoped regime, and rapid fall in Fermi velocity at low doping, are both mirrored by photoemission experiments – photoemission shows a collapse in the large Fermi surface on the overdoped side to disconnected nodal Fermi surface arcs on the underdoped side, the Fermi velocity of which rapidly decreases at low doping; (ii) the observation of quantum oscillations down to low magnetic fields where d-wave pairing is expected to strongly gap antinodal states and hence strongly attenuate quantum oscillations arising from any antinodal Fermi surface pockets; (iii) the quantum oscillation waveform is best explained by magnetic breakdown of a bilayer split nodal Fermi surface pocket.

An intriguing issue concerns the precise origin of the gapping of the electronic excitation spectrum at the antinodal region.  In part, this gapping may be expected to arise from the effects of the biaxial low-$\bf Q$ charge density wave state that provides a natural way of understanding a number of low temperature properties including, in particular, quantum oscillatory phenomena in the underdoped regions of interest here.  However, other major contributions to the formation of the antinodal gap are expected in addition to the low-Q charge density wave order emphasised here~\cite{lee1}.

\section{Acknowledgements}
SES acknowledges support from the Royal Society. NH acknowledges support from the DOE BES project ``Science at 100 Tesla.'' We acknowledge useful discussions with D. A. Bonn, B. Keimer, S. Kivelson, P. A. Lee, M. LeTacon, and thank M. LeTacon for sharing unpublished data with us.


\begin{thebibliography}{999}
\bibitem{notemultiple} Although the linear magnetoresistance at low temperatures has been interpreted as evidence for multiple Fermi surface sections comprising opposite charge carriers~\cite{vignolle2}, this behaviour has previously been observed in materials families in which the electronic structure consists of a single Fermi surface section including the overdoped cuprate superconductor Tl$_2$Ba$_2$CuO$_{6+\rm x}$ and various organic conductors, and is not necessarily related to multiple Fermi surface sections containing opposite carriers.
\bibitem{cp} The amplitude of oscillations in the heat capacity are yielded by the second derivative of the thermodynamic potential: $\frac{\tilde{C}}{T}= -6n_i\gamma_i \sum_{i=1}^{\infty} R^{''}_{T}R_{\rm S}R_{\rm D}R_{\rm{MB}} \cos\big(\frac{2\pi F_i}{B}\big)$, where $i$ labels each pocket, $n_i$ represents the number of times the $i^{\rm {th}}$ pocket occurs in the original Brillouin zone, $\gamma_i$ is the Sommerfeld coefficient associated with the $i^{\rm {th}}$ pocket, and $R_{T}$, $R_{\rm S}$, $R_{\rm D}$, and $R_{\rm{MB}}$ are the temperature, spin, Dingle, and magnetic breakdown damping factors respectively~\cite{shoenberg2}.
\bibitem{bondingnote} It has yet to be understood whether different wavevectors nest the bonding and antibonding bands independently in the bilayer system (giving rise to a separate charge modulation for that band) or whether a single period modulation prevails for both bonding and antibonding bands.
\end{thebibliography}
\end{document}